\documentclass[12pt,a4paper]{panl}
\usepackage{cite}
\usepackage{wrapfig}
\usepackage{graphicx}
\usepackage{amssymb}
\usepackage{amsfonts}
\usepackage{amsmath}
\usepackage{longtable}
\usepackage{rotating}
\usepackage{lscape}
\usepackage{epsfig}
\usepackage{multirow}
\usepackage{xcolor}

\originalTeX
\begin{document}

\title{Strongly intensive quantities for rapidity correlations of multiplicities}
\maketitle
\authors{E.\,Andronov$^{a}$\footnote{E-mail: e.v.andronov@spbu.ru}}
\setcounter{footnote}{0}
\authors{
Е.В.\,Андронов$^{a}$\footnote{E-mail: e.v.andronov@spbu.ru(русский вариант)}}
\from{$^{a}$\,Saint-Petersburg State University}
\from{$^{a}$\,Санкт-Петербургский Государственный Университет}

\begin{abstract}
Исследования фазовой диаграммы сильно взаимодействующей материи, образующейся при ядерных столкновениях, обычно проводятся с использованием пособытийных флуктуаций. Хорошо известный способ разделить статистические и динамические флуктуации заключается в построении специальных наблюдаемых величин, называемых сильно-интенсивными, которые свободны от тривиальных флуктуаций объема. В рамках модели цветных струн поведение сильно-интенсивной величины второго порядка $\Sigma$ полностью определяется двухчастичной корреляционной функцией от одной струны и механизмом слияния струн.\\
В этой статье мы исследуем сильно-интенсивную величину третьего порядка для быстротных корреляций вперед-назад и тестируем ее поведение в рамках модели PYTHIA8.

\vspace{0.2cm}

Studies of the phase diagram of strongly interacting matter created in nuclear collisions are typically carried out using event-by-event fluctuations. Well-known way to disentangle statistical and dynamical fluctuations is to construct special observables named strongly intensive which are free from trivial volume fluctuations. Within the color string model behavior of the second-order strongly intensive quantity $\Sigma$ is completely determined by the two-particle correlation function from a single string and the string fusion mechanism.\\
In this paper, we analyze third-order strongly intensive observable for forward-backward rapidity correlations and test its behavior within the PYTHIA8 model. 
\end{abstract}
\vspace*{6pt}


\label{sec:intro}
\section*{Introduction}

Over the past few decades, there has been a significant effort to study a special state of matter known as quark-gluon plasma (QGP)~\cite{Shuryak:1977ut} and to detect critical phenomena associated with phase transitions~\cite{Fodor:2004nz}. One of the main methods used in these studies is analyzing the statistical moments of various observables, such as particle multiplicities~\cite{ALICE:2021hkc} (for example, for particles of a certain species within a limited kinematic range), transverse momenta~\cite{ALICE:2024apz}, and net-charges~\cite{NA49:2012ebu,STAR:2025zdq}, as well as any combinations of these quantities when considering joint fluctuations~\cite{NA61SHINE:2020cxu,NA61SHINE:2015uhh,ALICE:2025mkk}. The idea behind this approach is that near the hypothetical critical point, large fluctuations may occur due to the divergence of the correlation length. For this goal it is important to disentangle statistical and dynamical fluctuations. A way to do this elegantly is to consider special type of observables called strongly intensive quantities~\cite{Gazdzicki:1992ri,Mrowczynski:1999un,Gorenstein:2011vq,Gazdzicki:2013ana,Broniowski:2017tjq}. These quantities are independent of the volume and volume fluctuations in simple baseline models by construction. It is in contrary to ratios of cumulants used in~\cite{STAR:2025zdq}, where additional data-driven techniques with limited applicability, such as the centrality-bin-width correction, are applied to suppress the contribution from volume fluctuations.

Similarly, forward-backward rapidity correlations of multiplicities $N_{F}$ and $N_{B}$ quantified by the correlation coefficient~\cite{Capella:1983dv}:
\begin{equation}
b_\mathrm{{corr}}[N_{F},N_{B}]= \frac{\langle N_{F}N_{B}\rangle - \langle N_{F}\rangle\langle N_{B}\rangle}{\langle N^2_{B}\rangle - \langle N_{B}\rangle^2}\,, \label{bcorrNFNB}
\end{equation}
are not free from volume fluctuations. In~\cite{Andronov:2015bqn} it was suggested to consider strongly intensive analogues of $b_\mathrm{{corr}}$:
\begin{eqnarray}
\Sigma[N_{F},N_{B}] &= &\frac{\langle N_{B}\rangle\omega[N_{F}] + \langle N_{F}\rangle\omega[N_{B}] - 2\left(\langle N_{F}N_{B}\rangle - \langle N_{F}\rangle\langle N_{B}\rangle\right)}{\langle N_{B}\rangle + \langle N_{F}\rangle}\,, \label{sigmaNFNB}\\
\Delta[N_{F},N_{B}] &= &\frac{\langle N_{B}\rangle\omega[N_{F}] - \langle N_{F}\rangle\omega[N_{B}]}{\langle N_{B}\rangle - \langle N_{F}\rangle}\,, \label{deltaNFNB}
\end{eqnarray}
where $\omega[A]=\frac{\langle A^{2}\rangle - \langle A\rangle^{2}}{\langle A\rangle}$ is the scaled variance of some observable $A$. Typically, correlations are studied as a function of the distance between the centers of the intervals, $\Delta y$, for windows that are placed symmetrically around the midrapidity point. Clearly, in this case $\Delta[N_{F},N_{B}]$ is asymptotically equal to $1$, therefore, for use of $\Delta[N_{F},N_{B}]$ it is more practical to consider asymmetric intervals.

In Fig.~\ref{fig01} we present the results obtained using the PYTHIA8.3/Angantyr event generator~\cite{Sjostrand:2019zhc,Bierlich:2016smv} for $10^{8}$ inelastic p+p interactions, $10^{7}$ min. bias O+O collisions and $10^{6}$ min. bias Xe+Xe collisions at a center-of-mass energy $\sqrt{s_{NN}} = 900$ GeV. The results are obtained for rapidity intervals of $\delta y=0.2$ width, which are placed symmetrically with respect to midrapidity. Clearly, the contribution from volume fluctuations is significantly suppressed in case of $\Sigma[N_{F},N_{B}]$ compared to $b_\mathrm{{corr}}[N_{F},N_{B}]$. 

\begin{figure}[!ht]
\begin{center}
\includegraphics[width=67mm]{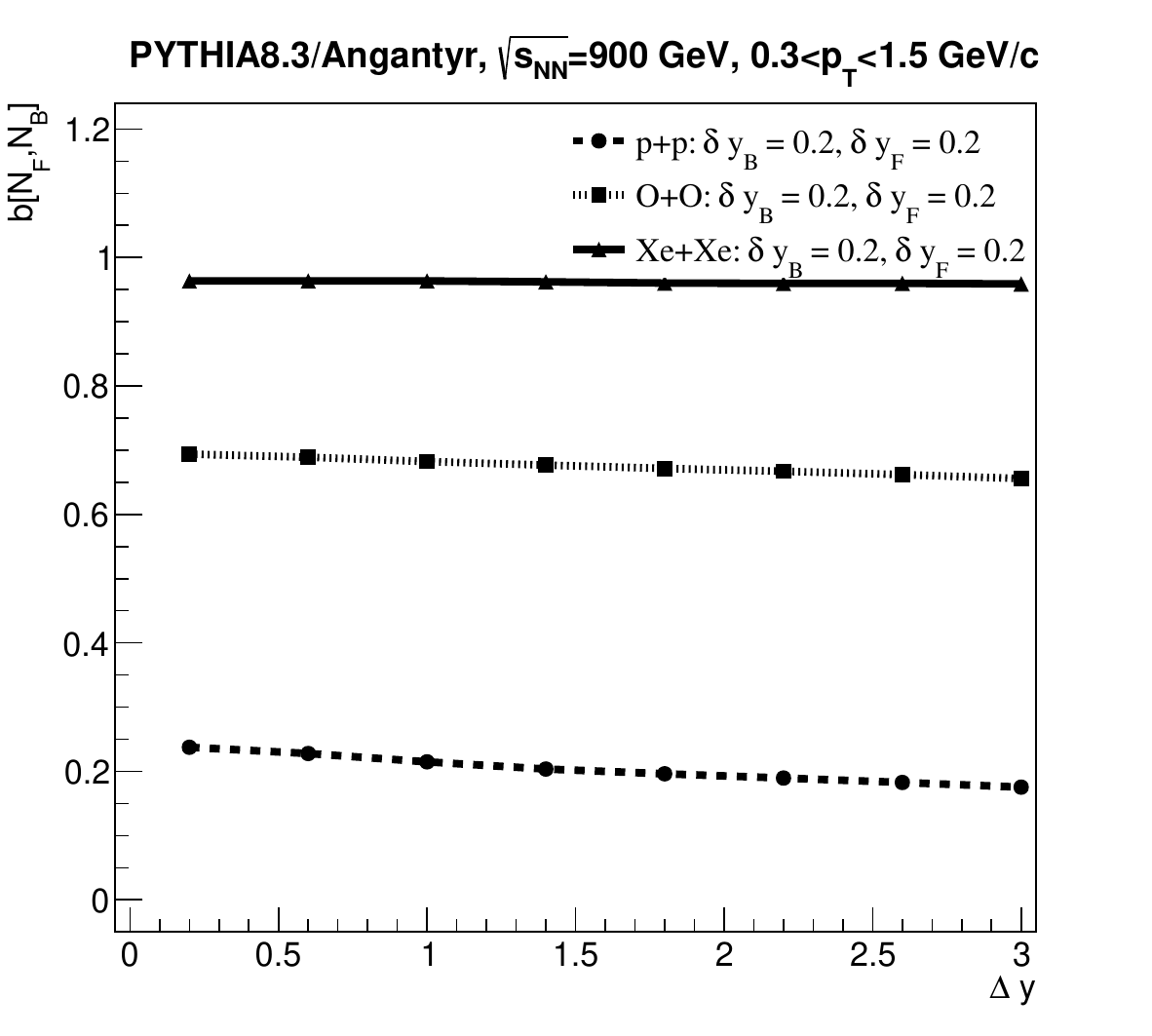}
\includegraphics[width=67mm]{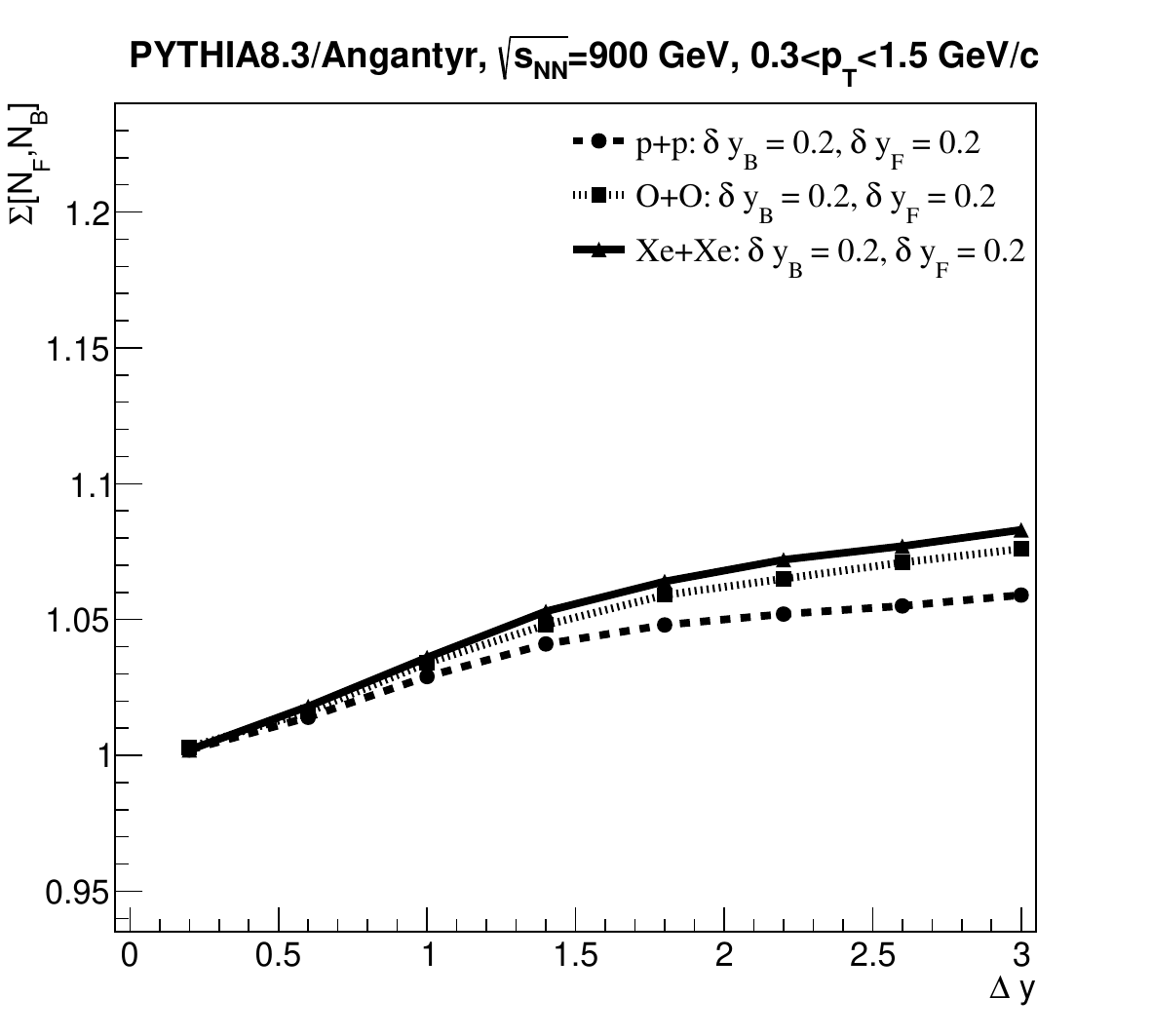}
\vspace{-3mm}
\caption{PYTHIA8.3/Angantyr predictions for the correlation coefficient, $b_\mathrm{{corr}}[N_{F},N_{B}]$, (left) and $\Sigma[N_{F},N_{B}]$, (right) calculated for particles with $0.3 < p_T < 3.0$ GeV/c in windows of $\delta y=0.2$ width produced in p+p, min. bias O+O and min. bias Xe+Xe collisions at $\sqrt{s_{NN}} = 900$ GeV as a function of distance between centers of intervals, $\Delta y$.}
\end{center}
\labelf{fig01}
\vspace{-5mm}
\end{figure}

This behavior is well understood in superposition models, such as the model of independent color strings, where
\begin{eqnarray}
\Sigma[N_{F},N_{B}] &= &\Sigma[\mu_{F},\mu_{B}]\,, \label{sigmaMuFMuB}\\
\Delta[N_{F},N_{B}] &= &\Delta[\mu_{F},\mu_{B}]\,. \label{deltaMuFMuB}
\end{eqnarray}
Here, $\mu_{B}$ and $\mu_{F}$ are multiplicities from a single string. The increase in $\Sigma[N_{F},N_{B}]$ with the increase of $\Delta y$ can be explained by the decrease of the two-particle correlation function from a single string. In~\cite{Andronov:2025dxn} this effect was modeled by introducing rapidity correlations by means of the Simulated annealing algorithm leading to description pf experimentally observed behavior~\cite{PHOBOS:2006mfc}. Equalities (\ref{sigmaMuFMuB}) and (\ref{deltaMuFMuB}) are slightly violated by string-string interactions as discussed in~\cite{Vechernin:2023ejn,Andronov:2023vnh,Prokhorova:2023hgq}: in regions where strings overlap the color field inside them rearranges leading to modifications in string fragmentation.  

\section*{Third-order strongly intensive quantities for forward-backward correlations} \label{sec2}

Following notations from~\cite{Broniowski:2017tjq} we denote joint cumulants of multiplicities $N_{F}$ and $N_{B}$ in two separated rapidity intervals as $P_{ij}$:
\begin{eqnarray}
P_{10} &=& \langle N_{F}\rangle,\\
P_{01} &=& \langle N_{B}\rangle,\\
P_{20} &=& \langle N_{F}^{2}\rangle-{\langle N_{F}\rangle}^{2},\\
P_{11} &=& \langle N_{B}N_{F}\rangle-{\langle N_{B}\rangle}\langle N_{F}\rangle,\\
P_{02} &=& \langle N_{B}^{2}\rangle-{\langle N_{B}\rangle}^{2},\\
P_{30} &=& \langle N_{F}^{3}\rangle-3{\langle N_{F}^{2}\rangle}\langle N_{F}\rangle+2{\langle N_{F}\rangle}^{3},\\
P_{21} &=& \langle N_{F}^{2}N_{B}\rangle-2{\langle N_{F}N_{B}\rangle}\langle N_{F}\rangle-\langle N_{F}^{2}\rangle\langle N_{B}\rangle+2{\langle N_{F}\rangle}^{2}\langle N_{B}\rangle,\\
P_{12} &=& \langle N_{B}^{2}N_{F}\rangle-2{\langle N_{F}N_{B}\rangle}\langle N_{B}\rangle-\langle N_{B}^{2}\rangle\langle N_{F}\rangle+2{\langle N_{B}\rangle}^{2}\langle N_{F}\rangle,\\
P_{03} &=& \langle N_{B}^{3}\rangle-3{\langle N_{B}^{2}\rangle}\langle N_{B}\rangle+2{\langle N_{B}\rangle}^{3}.
\end{eqnarray}
The same set of joint cumulants for the multiplicities $\mu_{F}$ and $\mu_{B}$ from a single string is denoted as $R_{ij}$. Cumulants in number of particle-emitting sources are denoted as $Q_{i}$. The relations between $P_{ij}$, $Q_{i}$ and $R_{ij}$ can be found in~\cite{Broniowski:2017tjq}. In order to construct strongly intensive observables one needs to combine $P_{ij}$ to exclude dependence on $Q_{i}$ in such a way that the functional dependence of the observable is the same at the level of $R_{ij}$:
\begin{eqnarray}
\Sigma[N_{F},N_{B}] &= & \frac{\frac{P_{01}P_{20}}{P_{10}}+\frac{P_{10}P_{02}}{P_{01}}-2P_{11}}{P_{10}+P_{01}}=\frac{\frac{R_{01}R_{20}}{R_{10}}+\frac{R_{10}R_{02}}{R_{01}}-2R_{11}}{R_{10}+R_{01}},\\
\Delta[N_{F},N_{B}] &= & \frac{\frac{P_{01}P_{20}}{P_{10}}-\frac{P_{10}P_{02}}{P_{01}}}{P_{01}-P_{10}}=\frac{\frac{R_{01}R_{20}}{R_{10}}-\frac{R_{10}R_{02}}{R_{01}}}{R_{01}-R_{10}},  
\end{eqnarray}
The expression for the third-order strongly intensive observable with the same property  was also found in~\cite{Broniowski:2017tjq}. We denote it as $\Gamma[N_{F},N_{B}]$ with the following normalization:
\begin{equation}
\Gamma[N_{F},N_{B}]=\frac{P_{10}P_{01}}{P_{01}^{2}-P_{10}^{2}}\left(\frac{P_{01}P_{30}}{P_{10}^{2}}-3\frac{P_{21}}{P_{10}}+3\frac{P_{12}}{P_{01}}-\frac{P_{10}P_{03}}{P_{01}^{2}}\right).
\end{equation}
Similarly to $\Delta[N_{F},N_{B}]$ this observable is best suited for asymmetric rapidity intervals.\\ 
Let us check some basic properties of strongly intensive observables in baseline models. First of all, in the model of independent particle production~\cite{Gorenstein:2011vq} one can define the probability for particle to be present in forward ($\frac{f}{f+b}$) or backward ($\frac{b}{f+b}$) interval, then
\begin{eqnarray}
    R_{01}&=&b,\\
    R_{10}&=&f,\\
    R_{02}&=&b-b^{2},\\
    R_{11}&=&-fb,\\
    R_{20}&=&f-f^{2},\\
    R_{03}&=&b-3b^{2}+2b^{3},\\
    R_{12}&=&-fb+2fb^{2},\\
    R_{21}&=&-fb+2f^{2}b,\\
    R_{30}&=&f-3f^{2}+2f^{3}.
\end{eqnarray}
Combining these cumulants we get baseline values:
\begin{equation}
    \Sigma[N_{F},N_{B}]=\Delta[N_{F},N_{B}]=\Gamma[N_{F},N_{B}]=1.
\end{equation}
The same baseline values are obtained in the superposition model if we assume that the probability distribution from a single string factorized into two Poisson distributions $P\left(\mu_{F},\mu_{B}\right)=P_{Pois}\left(\mu_{F}\right)P_{Pois}\left(\mu_{B}\right)$. This factorization assumption is expected to hold for $\Delta y\gg 1$. The assumption of the Poisson distribution is expected to be valid for small intervals $\delta y\ll 1$.

In this paper, we present a first look at $\Gamma[N_{F},N_{B}]$ behavior in the PYTHIA model. In Fig.~\ref{fig02}, left, results are presented for p+p and min. bias O+O collisions at $\sqrt{s_{NN}} = 900$ GeV for asymmetric rapidity intervals: backward interval is fixed at $y_{B}\in\left(-0.8,-0.75\right)$ and forward interval is moving from $y_{F}\in\left(-0.7,-0.6\right)$ to $y_{F}\in\left(1.5,1.6\right)$. We see that the property of strong intensity is present up to $\Delta y<1.8$. Possible breaking of strong intensity at larger $\Delta y$ due to string interactions would be addressed in future studies. In Fig.~\ref{fig02}, right, results are presented for p+p collisions at $\sqrt{s_{NN}} = 900$ GeV for asymmetric rapidity intervals: backward interval is fixed at $y_{B}\in\left(-0.8,-0.75\right)$ and forward intervals of different widths ($\delta y_{F}=0.1; 0.2; 0.4$) are moving. One can see that for all values of $\delta y_{F}$ $\Gamma[N_{F},N_{B}]$ increases showing only moderate dependence on the interval's width.

\begin{figure}[!ht]
\begin{center}
\includegraphics[width=67mm]{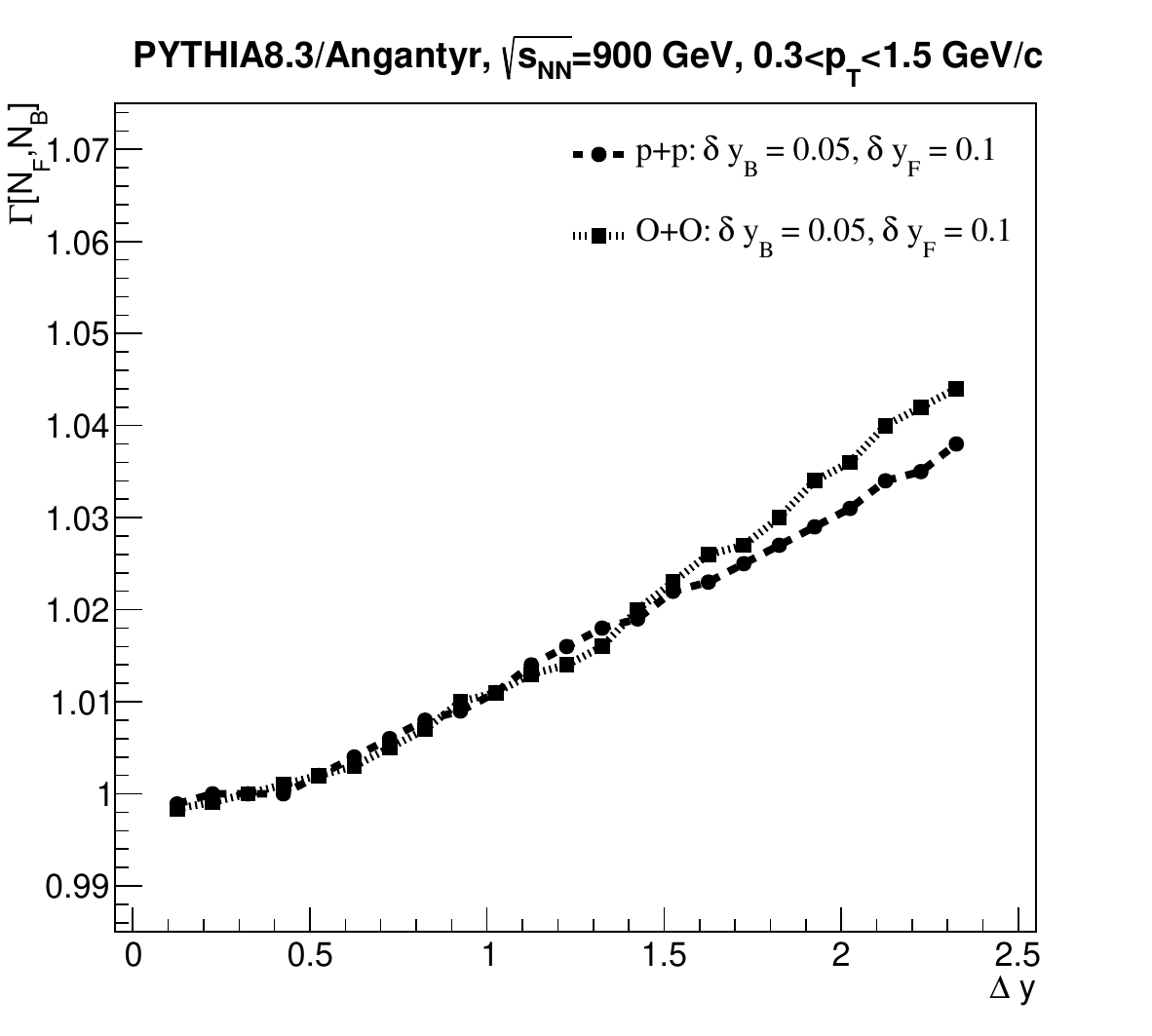}
\includegraphics[width=67mm]{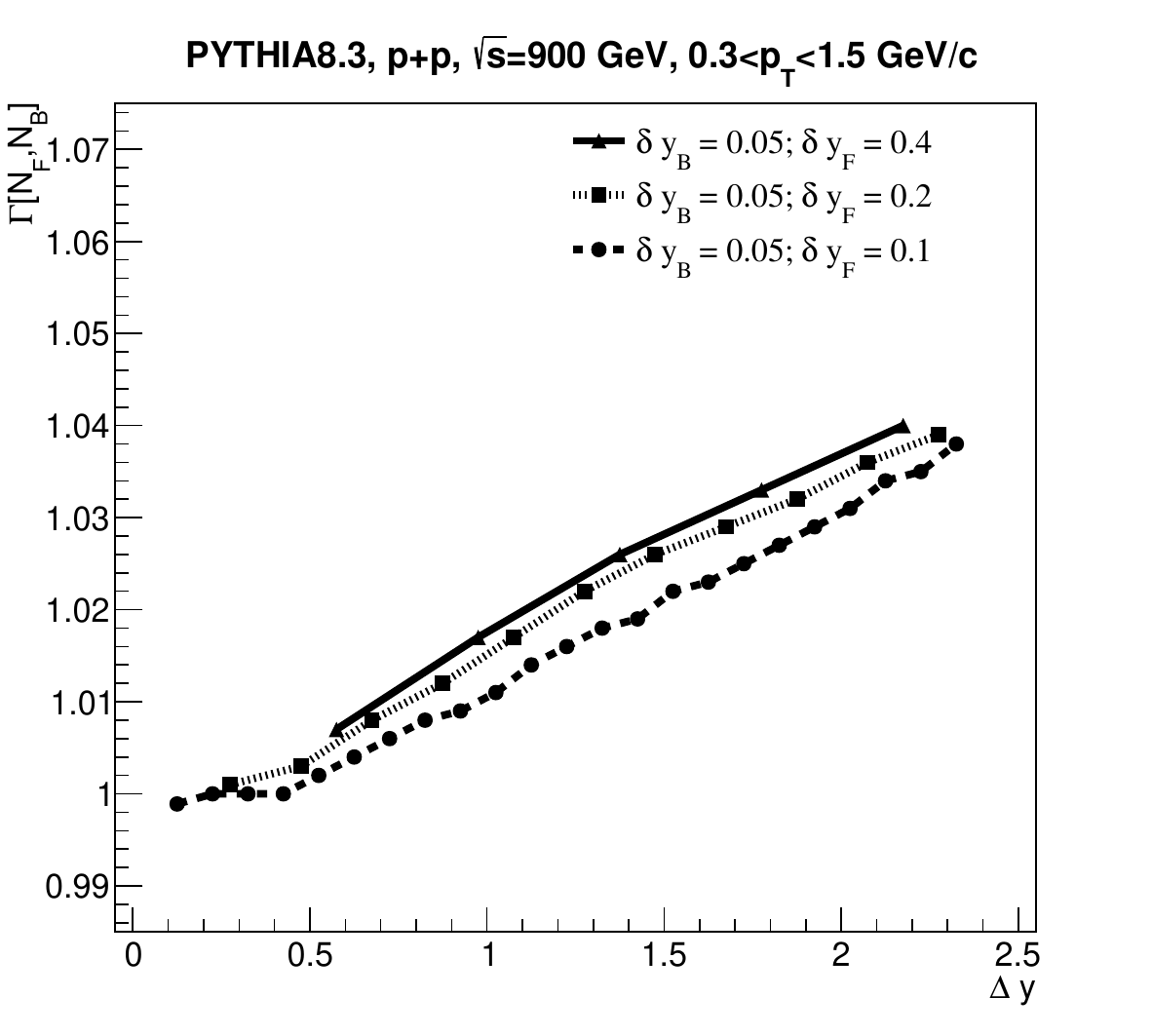}
\vspace{-3mm}
\caption{PYTHIA8.3/Angantyr predictions for $\Gamma[N_{F},N_{B}]$ calculated for particles with $0.3 < p_T < 3.0$ GeV/c as a function of distance between centers of intervals, $\Delta y$. Left: results for p+p and min. bias O+O collisions at $\sqrt{s_{NN}} = 900$ GeV for $\delta y_{B}=0.05$ and $\delta y_{B}=0.1$. Right: results for p+p collisions at $\sqrt{s_{NN}} = 900$ GeV for $\delta y_{B}=0.05$ and various $\delta y_{B}$.}
\end{center}
\labelf{fig02}
\vspace{-5mm}
\end{figure}

\section*{Summary}\label{sec6}
In this paper, we analyze the third-order strongly intensive quantity for forward-backward multiplicity correlations. The results obtained using the PYTHIA8 event generator suggest that new quantity $\Gamma$ behaves similarly to $\Sigma$ as the separation between rapidity intervals increases, and it has, indeed, the necessary property of strong intensity in the first approximation.

\section*{FUNDING}

The author acknowledges Saint-Petersburg State University for a research project 103821868.

\bibliographystyle{pepan}
\bibliography{pepan_biblio}

\end{document}